%% file: babar-pub-06045.tex
\documentclass[twocolumn,showpacs,prl,superscriptaddress,aps,nopreprintnumbers]{revtex4}

\usepackage{graphicx}
\usepackage{dcolumn}
\usepackage{epsfig}
\usepackage{amsmath}

\newcommand{\BaBarYear}    {06}
\newcommand{\BaBarNumber}  {045}
\newcommand{\SLACPubNumber}{11999}

\newcommand{\BaBarType}    {PUB}

\newcolumntype{q}{D{.}{.}{4.5}}

\input babarsym

\input Definitions

\renewcommand{\retapKst}{\ensuremath{4.1_{-0.9}^{+1.0} \pm 0.5}}
\renewcommand{\setapKst}{\ensuremath{5.6}}

\renewcommand{\retapKstz}{\ensuremath{3.8\pm1.1\pm0.5}}

\renewcommand{\setapKstz}{\ensuremath{4.3}}
\newcommand{\AetapKstz}{\ensuremath{-0.08 \pm 0.25 \pm 0.02}}

\renewcommand{\retapKstp}{\ensuremath{4.9^{+1.9}_{-1.7}\pm0.8}}
\renewcommand{\uletapKstp}{\ensuremath{7.9}}
\renewcommand{\setapKstp}{\ensuremath{3.6}}
\newcommand{\AetapKstp}{\ensuremath{0.30 ^{+0.33} _{-0.37} \pm 0.02}}

\renewcommand{\retaprhop}{\ensuremath{8.7^{+3.1}_{-2.8}{} ^{+2.3}_{-1.3}}}
\renewcommand{\uletaprhop}{\ensuremath{14}}
\renewcommand{\setaprhop}{\ensuremath{3.2}}
\renewcommand{\Aetaprhop}{\ensuremath{ -0.04 \pm 0.28 \pm 0.02}}

\renewcommand{\retaprhoz}{\ensuremath{0.4^{+1.2}_{-0.9}{}^{+1.6}_{-0.6}}}
\renewcommand{\uletaprhoz}{\ensuremath{3.7}}
\renewcommand{\setaprhoz}{\ensuremath{0.3}}

\renewcommand{\retapfz}{\ensuremath{0.1^{+0.6}_{-0.4}{}^{+0.9}_{-0.4}}}
\renewcommand{\uletapfz}{\ensuremath{1.5}}
\renewcommand{\setapfz}{\ensuremath{0.2}}

\begin{document}

\begin{flushleft}
\babar-\BaBarType-\BaBarYear/\BaBarNumber{} \\
SLAC-PUB-\SLACPubNumber 
\end{flushleft}

\title{
  \boldmath Observation of $B\ra\etapr\Kstar$ and Evidence for
  $\Bp\ra\etapr\rhop$ 
}

\input authors_jun2006.tex

\date{\today}

\begin{abstract}
  We present an observation of \etapKst.  The data sample corresponds
  to 232 million \BB pairs collected with the \babar{} detector at the
  PEP-II asymmetric-energy $B$ Factory at SLAC.  We measure the
  branching fractions (in units of $10^{-6}$) $\BetapKstz =
  \retapKstz$ and $\BetapKstp = \retapKstp$, where the first error is
  statistical and the second systematic.  A simultaneous fit results
  in the observation of \etapKst with $\mathcal{B}(B\ra\etapr\Kstar) =
  \retapKst$.  We also search for \etaprho and \fetapfz with results
  and 90\% confidence level upper limits $\Betaprhop = \retaprhop$
  ($<\uletaprhop$), $\Betaprhoz < \uletaprhoz$, and $\Betapfz <
  \uletapfz$.  Charge asymmetries in the channels with significant
  yields are consistent with zero.
\end{abstract}


\maketitle
%

Decays of $B$ mesons involving the flavor-changing neutral current
transition $b\to s$ are an important place to search for evidence of
physics beyond the Standard Model.  A comparison of the amplitude
$\sin 2\beta$ of time-dependent \CP{} violation in the neutral \CP{}
eigenstates $\jpsi\KS$ and $\etapr \KS$ provides one of the most
sensitive tests~\cite{Aubert:2005aa}. In order to unambiguously
interpret the time-dependent \CP{} violation measurement in $\etapr
\KS$ it is important to understand the full set of underlying
amplitudes by making measurements of branching fractions in the
$\etapr\Kst$ decays.

In \B decays to final states comprising $\etaprp K^{(*)}$ the final
states $\etapr\Kst$ and $\eta K$ are suppressed, and the final states
$\etapr K$ and $\eta\Kst$ are enhanced.  Two explanations of the
experimentally observed pattern differ substantially in the details of
the suppression for
$\B\ra\etapr\Kst$~\cite{Beneke:2002jn,Lipkin:1990us}.  From previous
experimental data and flavor SU(3) arguments it is expected that the
branching fractions for $\B\ra\etapr\Kst$ are less than
$10^{-5}$~\cite{Chiang:2003pm}.  The related decays \etaprho\ occur
via CKM suppressed tree diagrams and are expected to be small.
Theoretical approaches using QCD factorization~\cite{Beneke:2003zv}
and perturbative QCD~\cite{Liu:2005mm} predict branching fractions for
\etaprhop\ of $6$--$9\timesix$ and for \etaprhoz\ of
$0.5$--$2\times10^{-7}$.

In this Letter, we present searches for $B\ra\etapr\Kstar$,
$B\ra\etapr\rho$ and $B^0\ra\etapr f_0(980)(f_0\ra\pi^+\pi^-)$, which
shares the same final state as \etaprhoz.  Throughout this Letter,
charge conjugation is implied.  Results are obtained from unbinned,
extended maximum likelihood (ML) fits to data collected with the
\babar\ detector at the PEP-II asymmetric $e^+e^-$ collider located at
the Stanford Linear Accelerator Center.  The \babar\ detector and
relevant details specific to this analysis are described
elsewhere~\cite{bbr:NIM,Aubert:2004r12}.  The analysis uses 211
fb$^{-1}$ of data recorded at the \FourS resonance, corresponding to
232 million \BB{} pairs, and closely follows the approach described in
Ref.~\cite{Aubert:2004r12}.

We select $\etapr$, $K^*$, $\rho$, $\eta$, $\KS$ and $\pi^0$
candidates through the decays $\etapr\ra\eta\pi^+\pi^-$
($\etapr_{\eta\pi\pi}$), $\etapr\ra\rho^0\gamma$
($\etapr_{\rho\gamma}$), $\Kstarz\ra K^+\pi^-$, $\Kstarp\ra
K^0_S\pi^+$ ($\Kstarp_{K^0\pi^+}$), $\Kstarp\ra K^+\pi^0$
($\Kstarp_{K^+\pi^0}$), $\rho^0$ (and $f_0$) $\ra\pi^+\pi^-$,
$\rho^+\ra\pi^+\pi^0$, $\eta\ra\gamma\gamma$, $K^0_S\ra\pi^+\pi^-$ and
$\pi^0\ra\gamma\gamma$.  We impose the following requirements on
candidate invariant masses, in \mevcc:
$910<(m_{\eta\pi\pi},m_{\rho\gamma})<1000$ for $\etapr$,
$755<m_{K\pi}<1035$ for the $K^*$, $510< m_{\pi\pi^0}<1070$ for
$\rho^+$ and $510<m_{\pi\pi}<1060$ for $\rho^0$ ($f_0$),
$490<m_{\gamma\gamma}<600$ for $\eta$, $486<m_{\pi\pi}<510$ for
$K_S^0$ and $120<m_{\gamma\gamma}<150$ for $\pi^0$.  For the masses of
the \etapr, \Kst{} and $\rho$, which will be included as observables
in the ML fit described below, the selection is wide enough to allow
for a parameterization of the background.  For $\KS$ candidates we
require a flight distance of at least three times its estimated
uncertainty.

We also use the helicity-frame decay angle $\theta_H$ of $\Kstar$,
$\rho$, and $f_0(980)$.  The helicity frame is defined as the vector
meson rest frame with polar axis along the direction of the boost from
the $B$ rest frame.  The angle $\theta_H$ is the angle between the
polar axis and the flight direction of the charged resonance daughter.
For $\Kstarz$ and $\rhoz$ the kaon candidate and the positively
charged pion, respectively, are used to define that angle.  We use
mode dependent selection criteria on $\cos\theta_H$, with the lower
bound between $-0.95$ and $-0.70$ and the upper bound of either 0.95
or 1.00.  Decay modes suffering from higher combinatoric background
due to low momentum pions have the tighter cuts applied. The helicity
has a $\cos^2\theta_H$ distribution for $K^*$ and $\rho$ signal events
and is flat for the $f_0(980)$.

All charged pion candidates are required to have particle
identification (PID) consistent with pions and inconsistent with
protons, kaons, and electrons.  No such requirement is made of $\KS$
daughters.  Charged kaon candidates are required to have PID
consistent with kaons and inconsistent with pions, protons and
electrons.

We form $B$ meson candidates by combining an $\etapr$ candidate with
either a $\Kstar$ or $\rho$ candidate.  $B$ meson candidates are
characterized kinematically by the energy substituted mass, $\mes =
(s/4 - \mathbf{p}_B^2)^{1/2}$ and the energy difference $\DeltaE = E_B
- {\sqrt{s}}/{2}$ where $(E_B,\pvec_B)$ is the four-momentum of the
$B$ candidate, expressed in the \FourS frame and $\sqrt{s}$ is the
\epem{} center of mass energy.  Signal events peak at zero for \DeltaE
and at the $B$ mass for \mes, with typical resolutions of 20 MeV and
3.0 \mevcc, respectively.  We require $5.25\le\mes\le 5.29$ \gevcc{}
for all modes, $-0.2\le\DeltaE\le 0.150$ GeV for modes where the
vector meson decay includes a neutral pion and $-0.2\le\DeltaE\le
0.125$ GeV otherwise.

Backgrounds arise primarily from random combinations of particles in
continuum $e^+e^-\ra \qqbar$ ($q=u,d,s,c$) events.  To reject these
events, we employ the angle $\theta_T$ in the \FourS frame between the
thrust axis of the $B$ candidate's daughters and that of the remaining
particles in the event.  Continuum events are produced well above
threshold, with a jet-like topology resulting in a distribution of $|
\cos \theta_T |$ that is sharply peaked near 1 for candidates formed
in such events.  Events containing true \FourS decays are produced
near threshold with particles distributed isotropically, resulting in
a uniform distribution of $| \cos \theta_T |$.  We require $| \cos
\theta_T |< 0.9$ for decays with $\etapepp$, and $| \cos \theta_T |<
0.75$ for the higher-background $\etaprg$ decays.  Due to large
backgrounds in $\etaprg$, we only use the $\etapepp$ decay in
reconstructing $B\to\etapr\rho/f_0(980)$.

Additional discrimination against continuum background occurs in the
ML fit and is provided by a Fisher discriminant, ${\cal F}$. This is a
linear combination of discriminating variables with weights chosen to
maximize the separation between signal and continuum background.
$\cal F$ contains the angles of the $B$ momentum and $B$ thrust axis
with respect to the beam axis, the $B$-flavor tagging
category~\cite{Aubert:2004zt}, and the zeroth and second angular
moment of the energy flow in the rest of the event with respect to the
B candidate thrust axis~\cite{Aubert:2004r12}.

After selection, events containing multiple $B$ candidates occur less
than 30\% of the time.  In such cases, we choose the $B$ candidate
with the $\etapr$ mass closest to the Particle Data Group (PDG)
value~\cite{PDG2004}.

We use Monte Carlo (MC) simulation \cite{bbr:geant} for an initial
survey of background from \BB{} events and to identify for detailed
study any decays that are not rejected by candidate selection.  The
remaining background is composed almost entirely of charmless resonant
$B$ decays, especially $B\ra\etapr K$.  We account for $B$ backgrounds
by including in the ML fit an additional component which models these
charmless, resonant decays.  Backgrounds arising from charmed $B$
decays have been studied and found to be negligible or accounted for
by our continuum background model.  Backgrounds from non-resonant $B$
decays have been found to be consistent with zero.

We determine yields and charge asymmetries (${\cal A}_{ch} = (n^+ -
n^-) / (n^+ + n^-)$) for each decay chain from a ML fit with the
observables \DeltaE, \mes, ${\cal F}$, $m_{\etapr}$, the mass of the
candidate vector meson $m_V$, and ${\cal H} \equiv \cos\theta_H$.  For
charged (neutral) $B$ decays, $n^\pm$ is defined as the number of
$B^\pm$ decays (final states with $K^\pm$).  For each event $i$ and
hypothesis $j$ (signal, continuum, \BB), we define the probability
density function (PDF) as a simple product of the individual
observable PDFs:
\begin{equation*}
{\cal P}^i_j = {\cal P}_j(\mes^i){\cal P}_j(\DeltaE^i){\cal P}_j({\cal F}^i)
{\cal P}_j(m_{\etapr}^i){\cal P}_j(m_V^i){\cal P}_j({\cal H}^i).
\end{equation*}
For the $\etapr\pi^+\pi^-$ final state, a fourth hypothesis is added to account 
explicitly for a possible $\etapr f_0$ signal.

The total likelihood function is then given by
\begin{equation*}
{\cal L} = \frac{\exp (-\sum_j n_j)}{N!}\prod_{i}^N (\sum_j n_j{\cal P}^i_j),
\end{equation*}
where $N$ is the number of events in the sample and $n_j$ is the yield
of events of hypothesis $j$ to be found by maximizing ${\cal L}$.  In
addition to the yields and ${\cal A}_{ch}$ for each hypothesis,
parameters describing the continuum PDFs are also allowed to vary (see
below).

We parameterize the PDFs for peaking observables with either a single
or asymmetric Gaussian, sum of two Gaussians, or a Breit-Wigner as
required. Slowly varying observables are described by low degree
polynomials or phase-space motivated functions~\cite{Aubert:2004r12}.
Several PDFs require linear combinations of peaking and non-peaking
shapes.  We parameterize the $f_0(980)$ mass and width using measured
values~\cite{Aitala:2000xt}.

\input restab

For the signal and \BB{} background components we determine the PDF
parameters from simulation.  Control samples with topologies similar
to our signal (e.g., $\B^-\ra D^0\pi^-$) are used to verify and adjust
simulated resolutions~\cite{Aubert:2004r12}.  For the continuum
background we obtain initial PDF parameters from data excluding the
\DeltaE and \mes signal region (sideband).  We further refine the
continuum PDFs by letting as many parameters as feasible vary in the
fit to the full data.  The final fitted continuum background PDF
parameters are found to be in close agreement with their initial
values.

We apply several tests to the fitting procedure for validation before
implementing it on the data.  In particular, we evaluate any possible
bias in our event yields due to our neglect of small correlations
between the observables, which our PDFs ignore by construction.  We
determine the bias by fitting ensembles of simulated continuum
experiments generated from the PDF into which we embed the expected
number of signal and \BB{} background events randomly taken from
samples of fully simulated MC events.  Measured correlations in the
sideband data (pure \qqbar) are found to be small.  The measured
biases for each decay chain are given in Table~\ref{tab:results}.

We compute the branching fraction for each decay by subtracting the
fit bias from the measured yield and dividing the result by the
efficiency (determined from simulation and ancillary studies), the
product of the daughter branching fractions, and the number of
produced \BB pairs.  We assume equal decay rates of the \FourS to
\BpBm and \BzBzb.  In Table~\ref{tab:results} we show for each decay
the measured branching fraction, event yield, efficiency and daughter
branching fraction as well as ${\cal A}_{ch}$.

Measurements for separate decay chains are combined by adding the
values of $-2\ln{\cal L}$ as functions of branching fraction, taking
appropriate account of correlated and uncorrelated systematic
uncertainties (described below) \cite{Aubert:2004r12}.  The
significance is taken as the square root of the difference between the
value of $-2\ln{\cal L}$ (including systematics) for zero signal and
the minimum.  For modes where the combined significance is less than 4
standard deviations, we quote 90\% confidence level (C.L.) upper
limits.  We compute these as the branching fraction below which lies
90\% of the total likelihood integral in the positive branching
fraction domain.

\begin{figure}[!htb]
 \includegraphics[width=0.5\textwidth]{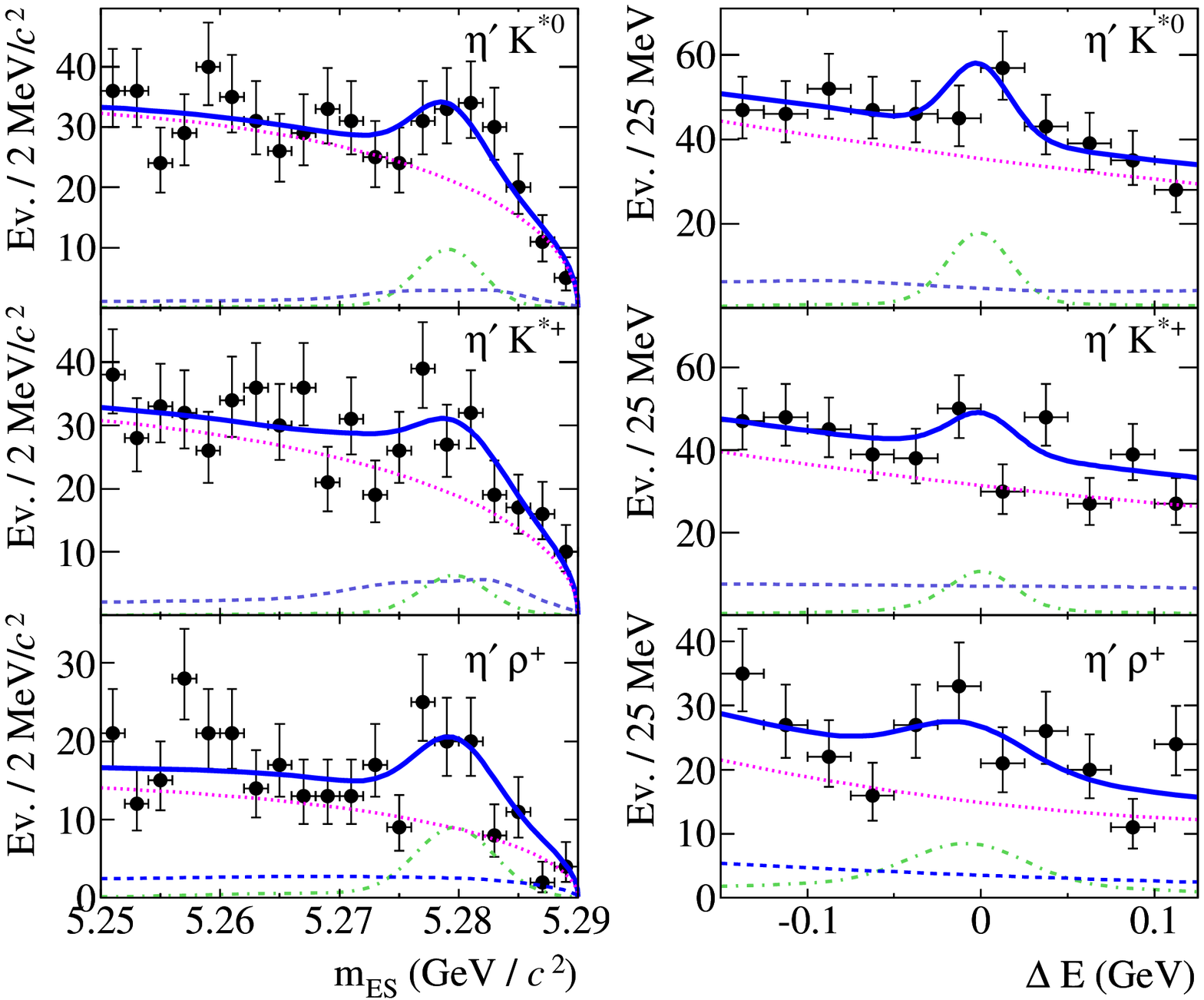}
\vspace{-0.75cm}
 \caption{\label{fig:proj} (Color online)
   $B$ candidate \mes (\emph{left}) and \DeltaE (\emph{right})
   projections obtained with a cut on the likelihood (see text) for
   \etapKstz (\emph{top}), \etapKstp (\emph{middle}) and \etaprhop
   (\emph{bottom}).  Submodes have been combined.  The data are
   represented by points with uncertainties, full fit functions by
   solid curves, \BB{} background by dashed, continuum by dotted and
   signal by dot-dashed curves. Depending on the decay, the plots
   contain 63 -- 85\% of all signal events.
}
\end{figure}

\begin{figure}[htb]
  \includegraphics[angle=0,width=0.5\textwidth]{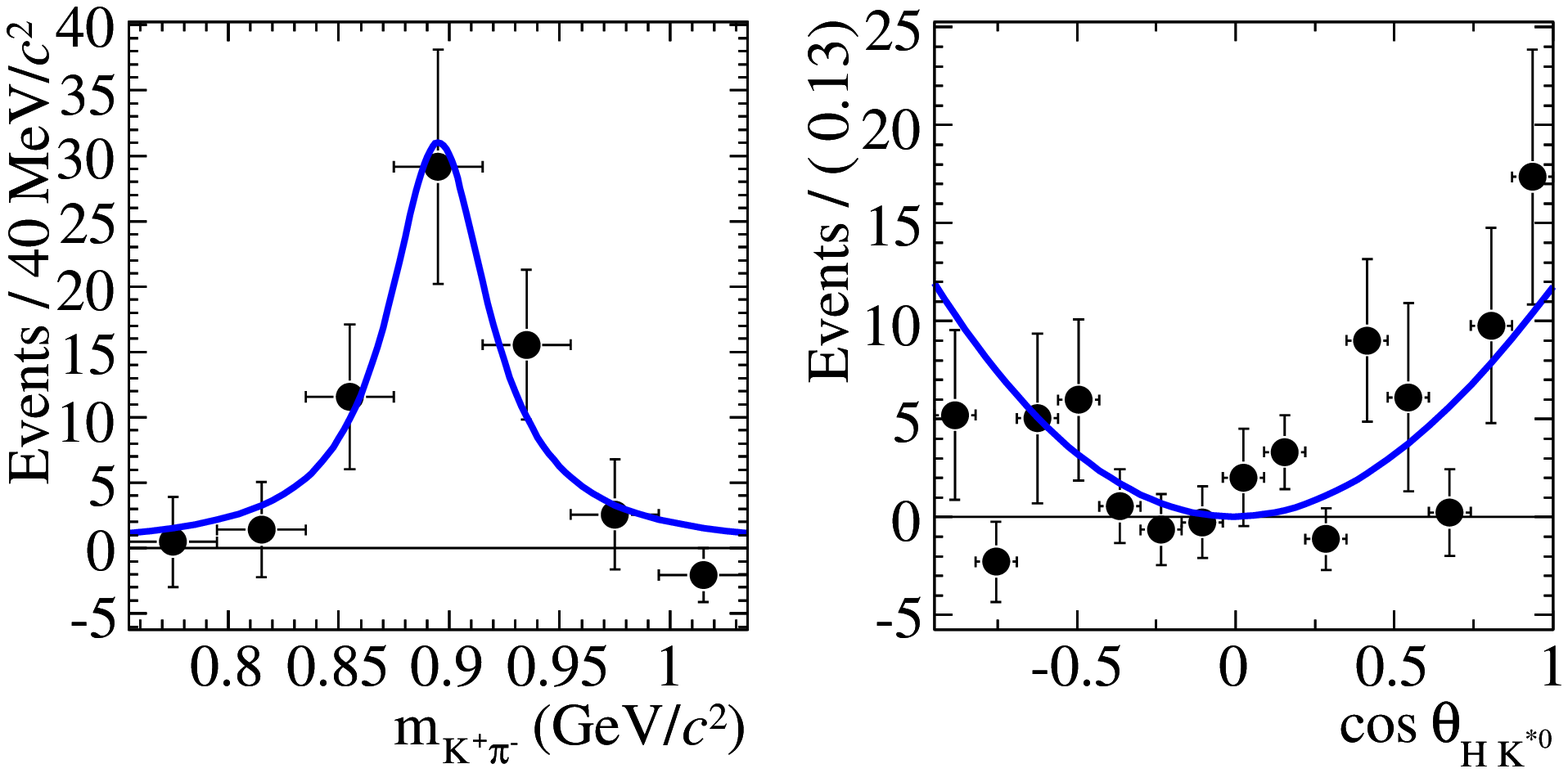}
\vspace{-0.75cm}
  \caption{\label{fig:kstarz_splots} (Color online)
    Distributions of the $K\pi$ mass (\emph{left}) and helicity
    (\emph{right}) for the decay $\etapKstz$. Points with error bars:
    data, background subtracted with the sPlot technique, solid curve:
    signal PDF.  }
\end{figure}

For modes with evidence of a signal, we show in Fig.~\ref{fig:proj}
projections onto \mes and \DeltaE of subsamples (containing 63 -- 85\%
of all signal events) enriched by a requirement on the ratio of the
signal likelihood to the total likelihood.  The likelihood is computed
excluding the plotted variable.  Figure~\ref{fig:kstarz_splots} shows
background-subtracted distributions of the \Kstarz mass and helicity
obtained with the sPlot technique described in~\cite{pivk:splots}.
These plots illustrate that the $K\pi$ signal we observe is consistent
with the $\Kstar(892)$ and is polarized as one would expect in a
pseudoscalar-vector $B$ decay.

Systematic uncertainties in this analysis are dominated by our
knowledge of signal and \BB{} background PDF modeling, along with the
fit bias and the efficiencies of the track and neutral particle
selections.  Uncertainty due to continuum PDF modeling is largely
incorporated into the statistical uncertainty since most continuum
background parameters are allowed to vary in the fit.  Uncertainties
in the signal PDF parameters are estimated from comparisons between
data and MC in control samples.  Varying the signal PDF parameters
within these errors results in a mode dependent variation in signal
yield of between 0.1 and 1.6 events.

The uncertainty in the fit bias is taken to be half of the correction.
We estimate the uncertainty from \BB{} modeling by taking half of the
difference between the signal yield fitted with and without the \BB
component ($0.2$ to $10$ events).  The uncertainty due to non-resonant
\BB background is estimated by taking half the difference between the
signal yield in the nominal fit and in a fit in which a non-resonant
background component has been added (0.7 to 4.8 events).
Uncertainties in reconstruction efficiency are determined from
supplementary studies of control samples.  These include 0.8\% per
charged track (excluding daughters of the $\KS$), 1.5\% per photon,
and 1.9\% for a $\KS$.  The systematic uncertainty in the number of
\BB pairs is 1.1\%~\cite{Aubert:2002hc}.  Published
data~\cite{PDG2004} provide the uncertainties in the $B$-daughter
product branching fractions (3.4\%).  Uncertainties in the event
selection efficiency are $0.5$--$3\%$ for the requirement on $\cos
\theta_T$.

We assign a systematic uncertainty on ${\cal A}_{ch}$ of 0.02, based
on studies of inclusive samples of kaons and $B$ decays.  This is due
primarily to asymmetries in charged kaon identification and slow pion
reconstruction.

We present measurements for the decays
$B^{+,0}\to\etapr\Kstar{}^{+,0}$ and $\Bp\ra\etapr\rho^+$.  They allow
the level of suppression of these decays, with respect to the enhanced
$\etapr K$ and $\eta\Kst$, to be determined.  A simultaneous fit of
all charged and neutral $\etapr\Kstar$ submodes results in the
observation of $\etapKst$ with a total significance of $\setapKst
\sigma$, including systematics, as shown in Table~\ref{tab:results}.
The measurements place constraints on possible enhanced flavor-singlet
contributions to these decays \cite{Beneke:2002jn,Chiang:2001ir}.
These results are consistent with previous upper limits, where they
existed.  In all cases, predictions based on SU(3) flavor
symmetry~\cite{Chiang:2003pm}, QCD factorization~\cite{Beneke:2003zv}
and perturbative QCD \cite{Liu:2005mm} are in excellent agreement with
our measured central values. Values of ${\cal A}_{ch}$ are consistent
with zero in all channels.

\input acknow_PRL

\bibliography{etapV}

\end{document}

%% file: Definitions.tex
\newcommand{\calB}{\ensuremath{{\cal B}}}

\newcommand{\bfemsix}{\ensuremath{\calB(10^{-6})}}
\newcommand{\timesix}{\ensuremath{\times10^{-6}}}
\newcommand{\pvec}{{\bf p}}

\newcommand{\etapepp}{\ensuremath{\etapr_{\eta\pi\pi}}}
\newcommand{\etaprg}{\ensuremath{\etapr_{\rho\gamma}}}

\newcommand{\etaprp}{\ensuremath{\eta^{(\prime)}}}      
\newcommand{\Kst}{\ensuremath{K^*}}

   \newcommand{\rhop}{\ensuremath{\rho^+}}
   \newcommand{\rhoz}{\ensuremath{\rho^0}}

\newcommand{\retapKst}{\ensuremath{xx^{+xx}_{-xx}\pm xx}}
\newcommand{\setapKst}{\ensuremath{xx}}

\newcommand{\fetapKstz}{\ensuremath{\etapr K^{*0}}}
\newcommand{\etapKstz}{\ensuremath{\Bz\ra\fetapKstz}}
\newcommand{\BetapKstz}{\ensuremath{\calB(\etapKstz)}}
\newcommand{\retapKstz}{\ensuremath{xx^{+xx}_{-xx}\pm xx}}

\newcommand{\setapKstz}{\ensuremath{xx}}
   \newcommand{\fetapeppKstz}{\ensuremath{\etapr_{\eta\pi\pi} K^{*0}}}

   \newcommand{\fetaprgKstz}{\ensuremath{\etapr_{\rho\gamma} K^{*0}}}
   
\newcommand{\etapKst}{\ensuremath{\B\ra\etapr K^{*}}\xspace}

\newcommand{\etapKstp}{\ensuremath{\Bp\ra\etapr K^{*+}}}
\newcommand{\BetapKstp}{\ensuremath{\calB(\etapKstp)}}
\newcommand{\retapKstp}{\ensuremath{xx^{+xx}_{-xx}\pm xx}}

\newcommand{\uletapKstp}{\ensuremath{xx}}

\newcommand{\setapKstp}{\ensuremath{xx}}

   \newcommand{\fetapeppKstpKspip}{\ensuremath{\etapr_{\eta\pi\pi}\Kstarp_{\Kz\pip}}}
   
   \newcommand{\fetapeppKstpKppiz}{\ensuremath{\etapr_{\eta\pi\pi}\Kstarp_{\Kp\piz}}}
   
   \newcommand{\fetaprgKstpKspip}{\ensuremath{\etapr_{\rho\gamma}\Kstarp_{\Kz\pip}}}
   
   \newcommand{\fetaprgKstpKppiz}{\ensuremath{\etapr_{\rho\gamma}\Kstarp_{\Kp\piz}}}
   
\newcommand{\fetaprho}{\ensuremath{\etapr\rho}\xspace}

\newcommand{\etaprho}{\ensuremath{\B\ra\fetaprho}\xspace}

\newcommand{\fetaprhop}{\ensuremath{\etapr\rho^+}}
\newcommand{\etaprhop}{\ensuremath{\Bp\ra\fetaprhop}}
\newcommand{\Betaprhop}{\ensuremath{\calB(\etaprhop)}}
\newcommand{\retaprhop}{\ensuremath{xx^{+xx}_{-xx}\pm xx}}

\newcommand{\uletaprhop}{\ensuremath{xx}}

\newcommand{\Aetaprhop}{\ensuremath{xx\pm xx \pm xx}}
\newcommand{\setaprhop}{\ensuremath{xx}}

\newcommand{\fetaprhoz}{\ensuremath{\etapr\rho^0}}
\newcommand{\etaprhoz}{\ensuremath{\Bz\ra\fetaprhoz}}
\newcommand{\Betaprhoz}{\ensuremath{\calB(\etaprhoz)}}
\newcommand{\retaprhoz}{\ensuremath{xx^{+xx}_{-xx}\pm xx}}

\newcommand{\uletaprhoz}{\ensuremath{xx}}

\newcommand{\setaprhoz}{\ensuremath{xx}}

\newcommand{\fetapfz}{\ensuremath{\etapr f_0(980)(f_0\ra\pi^+\pi^-)}\xspace}
\newcommand{\etapfz}{\ensuremath{\Bz\ra\fetapfz}\xspace}
\newcommand{\Betapfz}{\ensuremath{\calB(\etapfz)}\xspace}
\newcommand{\retapfz}{\ensuremath{xx^{+xx}_{-xx}\pm xx}\xspace}

\newcommand{\uletapfz}{\ensuremath{xx}\xspace}

\newcommand{\setapfz}{\ensuremath{xx}\xspace}

%% file: authors_jun2006.tex
%
\author{B.~Aubert}
\author{R.~Barate}
\author{M.~Bona}
\author{D.~Boutigny}
\author{F.~Couderc}
\author{Y.~Karyotakis}
\author{J.~P.~Lees}
\author{V.~Poireau}
\author{V.~Tisserand}
\author{A.~Zghiche}
\affiliation{Laboratoire de Physique des Particules, F-74941 Annecy-le-Vieux, France }
% ...
\collaboration{The \babar\ Collaboration}
\noaffiliation

%% file: restab.tex
\begin{table*}[btp]
\begin{center}
  \caption{Summary of results showing (from left): fitted signal yield $n$ before bias correction, 
    fit bias, detection efficiency $\varepsilon$, product
    daughter branching fraction $\prod\calB_i$~\cite{PDG2004}, 
    significance $S$ (including systematic uncertainties) in standard deviations,
    measured branching fraction $\calB$ and signal charge asymmetry ${\cal A}_{ch}$ for each mode.
    The values in parentheses are 90\% C.L. upper limits.  The result for $\etapfz$ includes
    the branching fraction for $f_0\ra\pi^+\pi^-$, which is not well known.  Results in bold face
    represent combined fits to multiple decay chains (when present).
  \label{tab:results}
}

\setlength{\extrarowheight}{4pt}
\begin{ruledtabular}
\begin{tabular}{lqqqqccc}
Mode & \multicolumn{1}{r}{$n$ (ev.)} & \multicolumn{1}{c}{Bias (ev.)} & 
\multicolumn{1}{c}{$\varepsilon (\%)$} & \multicolumn{1}{c}{ $\prod \calB_i (\%)$} & $S(\sigma)$ & $\bfemsix$ & ${\cal A}_{ch}$  \\
\hline
$\etapKst$                &&&&                                                    &\boldmath\setapKst  &\boldmath\retapKst                & \\ 
$\etapKstz$               &&&&                                                    &\boldmath\setapKstz &\boldmath\retapKstz               &\AetapKstz\\
$\quad \fetapeppKstz$     & 22.6^{+7.7}_{-6.7}  &+1.7\pm0.9   &19.0\pm1.2&11.6 &3.9                 &$4.1_{-1.3}^{+1.5}$               &\\
$\quad \fetaprgKstz$      & 35.1^{+14.2}_{-12.7}&+9.5\pm4.8   &16.9\pm1.1&19.7 &2.0                 &$3.3_{-1.6}^{+1.9}$               &\\
$\etapKstp$               &                     &&&                                  &\boldmath\setapKstp &\boldmath\retapKstp ($<\uletapKstp$) &\AetapKstp\\
$\quad \fetapeppKstpKspip$& 11.2^{+5.7}_{-4.5}  &+0.8\pm0.5   &18.0 \pm1.2&4.0 &3.2 &$6.2^{+3.4}_{-2.7}$ &\\
$\quad \fetaprgKstpKspip$ & 14.8^{+11.2}_{-9.7} &+2.9\pm1.5   &15.8 \pm1.1&6.8 &1.2 &$4.7^{+4.5}_{-3.9}$ &\\
$\quad \fetapeppKstpKppiz$&  5.2^{+5.4}_{-3.6}  &+1.0\pm0.5   &10.7 \pm0.6&5.8 &1.2 &$2.9^{+3.7}_{-2.6}$ &\\
$\quad \fetaprgKstpKppiz$ &  3.1^{+12.1}_{-9.6} &-2.3\pm1.3   &8.0  \pm0.5&9.8 &0.5 &$2.9^{+6.7}_{-5.4}$ &\\
$\etaprhoz$               & 14.9^{+10.6}_{-8.4} &+11.2\pm5.7  &22.8 \pm1.4&17.5&\setaprhoz&\retaprhoz ($<\uletaprhoz$) &\\
$\Bz\ra\etapr f_0(\ra \pi^+\pi^-)$    & -2.6^{+6.0}_{-4.0}  &-3.8\pm2.0   &25.4 \pm1.6&17.5&\setapfz&\retapfz ($<\uletapfz$)       &\\
$\etaprhop$               & 57.3^{+16.0}_{-14.7}&+11.5\pm5.8  &13.0 \pm1.0&17.5&\setaprhop&\retaprhop ($<\uletaprhop$) &\Aetaprhop\\
\end{tabular}
\end{ruledtabular}
\end{center}
\end{table*}


%% file: acknow_PRL.tex
We are grateful for the excellent luminosity and machine conditions
provided by our \pep2\ colleagues, 
and for the substantial dedicated effort from
the computing organizations that support \babar.
The collaborating institutions wish to thank 
SLAC for its support and kind hospitality. 
This work is supported by
DOE
and NSF (USA),
NSERC (Canada),
IHEP (China),
CEA and
CNRS-IN2P3
(France),
BMBF and DFG
(Germany),
INFN (Italy),
FOM (The Netherlands),
NFR (Norway),
MIST (Russia),
MEC (Spain), and
PPARC (United Kingdom). 
Individuals have received support from the
Marie Curie EIF (European Union) and
the A.~P.~Sloan Foundation.